# Degree of dimerization, effective polarizability of molecules and heat capacity of the saturated water vapor

*Viktor N. Makhlaichuk, Nikolay P. Malomuzh*

Dept. of Theoretical Physics, Odessa National University, 2 Dvoryanskaja str., Odessa, 65026, Ukraine, interaktiv@ukr.net

The work is devoted to the investigation of properties of water vapor. The main attention is focused on the physical nature of its effective polarizability and heat capacity at constant volume. We show that the specific temperature dependencies of these characteristics are mainly caused by thermal vibration excitations of water dimers. In connection with this we determine normal coordinates for the dimer vibrations. We investigate fluctuations of the dipole moments of dimers in details and consider their contributions to the dielectric permittivity of the water vapor. We also took into account the contribution of the interparticle interactions on the heat capacity. Analyzing the behavior of the effective polarizability and heat capacity we defined the temperature dependence of the dimer concentration on the coexistence vapor-liquid curve. The noticeable dimerization in the saturated water vapor takes place only for temperatures $T/T_c > 0.8$ ($T_c$ is the critical temperature). It is important to note that the successful reproduction of the effective polarizability and heat capacity can serve good test for the correct description of the dimer concentration in different approaches.

## 1. Introduction

There are many works [1-9] devoted to the dimerization of water vapor. The authors use different methods to investigate: 1) the determination of the dimer concentration $c_d$ with the help of the second virial coefficient [10-12]; 2) direct calculation [13-15], requiring the knowledge of spectra for vibration and rotational excitations of dimers and 3) the thermo-conductivity of water vapor. Unfortunately, the results differ essentially from each other.

From this follows that the analysis of new properties of water vapor can help us to choose the correct values of $c_d$. We consider that such important properties like 1) effective polarizability per molecule and 2) the heat capacity per molecule at constant volume are among them. In connection with this we would like to note the following.

The effective polarizabity of water molecules are determined by the standard equation:

$$\frac{\varepsilon-1}{\varepsilon+2}=\frac{4\pi}{3}n_w\alpha_{eff}, \quad n_w=\rho/m_w, \tag{1}$$

where $n_w$ and $\rho$ are the numerical and mass density of a water vapor. Near the triple point the water vapor is practically non dimerized and its effective polarizabity is reduced to

$$\alpha_{eff}=\alpha_e+\alpha_d, \quad \alpha_d=\frac{\vec{d}^2}{3k_BT}, \tag{2}$$

where $\alpha_e$ is the polarizability of electronic shell of water molecule and $\alpha_d=\frac{\vec{d}^2}{3k_BT}$ is the contribution caused by dipole moment of isolated water molecule. The analysis of the temperature dependence of $\alpha_{eff}$ for saturated water vapor in [15] shows that beginning from $t\geq 0.7$ we observe the essential deviations from (2). At that, the corresponding deviations increase with temperature.

It is clear that the general structure of $\alpha_{eff}$ is complicated. If we ignore the thermal excitations of dimers, we will get:

$$\alpha_{eff}=\frac{1}{1+c_d}\left[\alpha_e+\alpha_d+c_d\left(\alpha_e^{(d)}-\alpha_e\right)+c_d\left(\alpha_d^{(d)}-\alpha_d\right)\right], \tag{3}$$

where $\alpha_e^{(d)}$ and $\alpha_d^{(d)}=\frac{\vec{d}_d^2}{3k_BT}$ are the contributions to the effective polarizability caused by an electronic shell and dipole moment of a dimer. Additionally, we took into account that the total number of monomers and dimers in water vapor satisfy the equation:

$$n_m+n_d=\frac{1}{1+c_d}\rho/m_w. \tag{4}$$

However, such modification of the effective polarizability leads to the slight improvement of agreement with experimental results for $t\geq 0.7$. Therefore, the detailed consideration of the role of thermal excitations for dimers becomes the vital problem.

The same problems also take place for temperature dependence of the heat capacity $c_V$ at constant volume for the water vapor. Up to $t \leq 0.7$ it is quite satisfactory described by the sum:

$$c_V(t) = 6 + c_{int}(t), \qquad (5)$$

where $c_{int}(t)$ is the contribution caused by the interaction between water monomers (the heat capacity is measured in units of the Boltzmann constant $k_B$). However, the formula (5) becomes unsatisfactory for $t \geq 0.7$. Here the careful account of thermal excitation of dimers is necessary again.

The main goal of this paper is the detail study of the thermal excitations for water dimers and their manifestations in the effective polarizability of water molecules and the heat capacity at constant volume. We will consider 1) the specificity of normal coordinates for dimer; 2) thermal fluctuations of the dipole moment for dimer; 3) the contribution of thermal excitations to the free energy; 4) the generalization of $c_{int}(t)$ for the ensemble of monomers and dimers and 5) their joint influence on the behavior of the effective polarizability of a molecule and the heat capacity of water vapor. We will also test in details the temperature dependence for the concentration $c_d$ of dimers, obtained in [10-14].

## 2.  Normal coordinates for a water dimer vibrations

Two water molecules have 12 independent degrees of freedom (3 translational and 3 orientational ones per molecule). If these molecules form a dimer it is necessary 3 coordinates for the description of its center of mass and 3 angular variables for the description of its rotation as a whole. Thus, 6 degrees of freedom correspond to its vibration excitations.

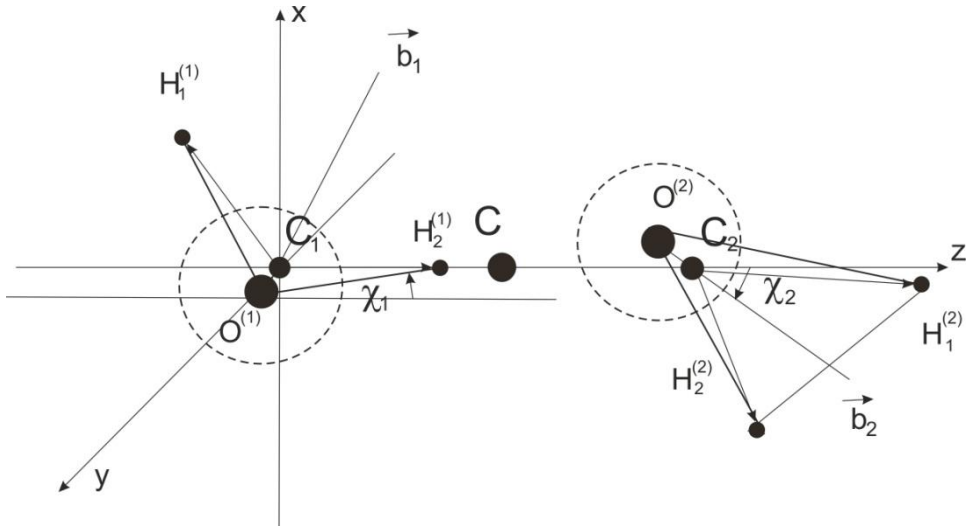

Fig.1. A water dimer in the molecular system of coordinates: 1) $C_1$, $C_2$ and $C$ denote the centers of masses of molecules and dimer correspondingly; 2) $H_1^{(i)}$, $H_2^{(i)}$ and $O^{(i)}$, $i = 1,2,$ denote the positions of hydrogens and oxygens; 3) $\vec{b}_1$, $\vec{b}_2$ are bisectors of molecules, directed along their dipole moments.

The choice of normal coordinates for the description of dimer vibrations discussed in [16]. Here we briefly discuss this question since the evident form of normal coordinates used in our calculations. The axes of the molecular coordinates system (MCS) are defined as: $z$-axis is directed along the line passing through the centers of mass of molecules $C_1$ and $C_2$ (see Fig.1), the $x$-axis lies in the plane formed by the bisectors of molecules in the equilibrium configuration of a dimer. We should take into account that vibrations of a dimer do not change 1) the impulse $\vec{P}$ for the center of mass of an isolated dimer and 2) its angular moment $\vec{M}$. Further we put: $\vec{P} = 0,\ \vec{M} = 0$.

Among the allowable following deviations of molecular coordinates from their equilibrium values are: 1) distance between points $C_1$ and $C_2$; 2) polar angles $\chi_1$ and $\chi_2$; 3) azimuth angles $\varphi_1$ and $\varphi_2$ as well as 3) rotation angles $\phi_1$ and $\phi_2$ around the bisectors.

The first normal coordinate is connected with the distance between the centers of masses of molecules: $\delta N_1 = \delta r_{C_1 C_2} \equiv \delta r$. In this case: $\delta z_1 = -\frac{1}{2}\delta r$ and $\delta z_2 = \frac{1}{2}\delta r$.

The angle variable $\delta N_2 = \delta \varphi$ will play the role of the second normal coordinate, if the angles $\varphi_1$ and $\varphi_2$ are connected with it by the relations: $\delta\varphi_1 = \delta\varphi$ and $\delta\varphi_2 = -a_2\delta\varphi$ , where $a_2$ follows from the equation $M_z = (I_{zz}^{(1)} - a_2 I_{zz}^{(2)})\delta\varphi = 0$ and is equal to: $a_2 = \dfrac{I_{zz}^{(1)}}{I_{zz}^{(2)}}$ ( $I_{zz}^{(1)}$ and $I_{zz}^{(2)}$ are the inertia moments of water molecules around $z$-axis).

The contributions to the kinetic energy caused by these normal coordinates are equal to:

$$
\begin{aligned}
&T_1 = \frac{1}{2}m_1\,\delta\dot{N}_1^2, \quad m_1 = \frac{1}{2}m_w \\
&T_2 = \frac{1}{2}m_2\,\delta\dot{N}_2^2, \quad m_2 = I_{zz}^{(1)} + a_2^2 I_{zz}^{(2)}, \quad a_2 = I_{zz}^{(1)} / I_{zz}^{(2)}
\end{aligned}
\tag{6}
$$

The third and fourth normal coordinates are connected with variations of angles $\delta\chi_1$ and $\delta\chi_2$. Let $\delta M_{y_1}^{(1)} = I_{y_1 y_1}^{(1)}\delta\dot{\chi}_1$ be the increment of the momentum of impulse for the left molecule. In accordance with the requirement $\vec{M} = 0$ it should be compensated by the rotation of the right molecule around the same axis $y_1$: $\delta M_{y_1}^{(2)} = I_{y_1 y_1}^{(2)}\delta\dot{\chi}_2 = -\delta M_{y_1}^{(1)}$. Since $I_{y_1 y_1}^{(2)}$ is approximately, on the order of magnitude, more than $I_{y_1 y_1}^{(1)}$ we can neglect by the variation of $\delta\dot{\chi}_2$. The analogous situation also takes place for rotations of the right molecule. Therefore these angular variables can be identified with two different normal coordinates: $\delta N_3 \approx \delta\chi_1$, $\delta N_4 \approx \delta\chi_2$. The corresponding contributions to the kinetic energy are equal to:

$$
\begin{aligned}
&T_3 = \frac{1}{2}m_3\delta\dot{N}_3^2, \quad m_3 = (1+a_3)I_{y_1 y_1}^{(1)}, \quad a_3 = I_{y_1 y_1}^{(1)} / I_{y_1 y_1}^{(2)} < 0.1, \\
&T_4 = \frac{1}{2}m_4\delta\dot{N}_4^2, \quad m_4 = \left(1+a_4\right)I_{y_2 y_2}^{(2)}, \quad a_4 = I_{y_2 y_2}^{(2)} / I_{y_2 y_2}^{(1)} < 0.1.
\end{aligned}
\tag{7}
$$

Because of the same causes the fifth and sixth normal coordinates can be identified with the rotation angles around the bisectors: $\delta N_5 = \delta \phi_1$ and $\delta N_6 = \delta \phi_2$. In this case:

$$T_5 = \frac{1}{2} m_5 \delta \dot{N}_5^2, \quad m_5 = (1 + a_5) I_{b_1 b_1}^{(1)}, \quad a_5 = I_{b_1 b_1}^{(1)} / I_{b_1 b_1}^{(2)} < 0.1. \tag{8}$$

$$T_6 = \frac{1}{2} m_6 \delta \dot{N}_6^2, \quad m_6 = (1 + a_6) I_{b_2 b_2}^{(2)}, \quad a_6 = I_{b_2 b_2}^{(2)} / I_{b_2 b_2}^{(1)} < 0.1. \tag{9}$$

One can verify that the inertia moments entering to (6) − (9) take the following values:

$$I_{zz}^{(1)} = mr_0^2 \left( \sin^2(\delta + \chi_1) + \sin^2 \chi_1 - \frac{30}{81} \sin(\delta + \chi_1) \sin \chi_1 \right), \tag{10}$$

$$I_{y_1 y_1}^{(1)} = 2mr_0^2, \ I_{y_2 y_2}^{(2)} = \frac{144}{81} mr_0^2 \cos^2 \delta / 2, \ I_{b_1 b_1}^{(1)} = I_{b_2 b_2}^{(2)} = 2mr_0^2 \sin^2 \delta / 2, \tag{11}$$

Here $m$ is the hydrogen mass and $r_0$ is the distance between oxygen and hydrogen in a water molecule. Besides, we use below: $m_w = 18m$, $r_{C_1 C_2} \approx r_{O_1 O_2} = 2.98$ Å, $r_0 \approx \frac{1}{3} r_{C_1 C_2}$, $\chi_1 \approx 2.44^0$, $\chi_2 \approx 22.19^0$ and $\delta = 109^0$ [17].

Values of the coefficients $a_{i_1}$ are placed in the Table 1.

### 3. Mean-square fluctuations of the dipole moment for water dimer

It follows from the symmetry reasons that fluctuations of the dipole moment for a water dimer are only caused by excitations of the normal coordinates: $\delta N_2$, $\delta N_3$ and $\delta N_4$. Let us consider consequently all these cases.

The increment of the dipole moment of a dimer corresponding to rotations of water molecules around $z$ − axis is equal to

$$\delta \vec{d}_D = \delta \vec{d}_1 + \delta \vec{d}_2, \tag{12}$$

where

$$\delta\vec{d}_1 = \vec{j}\,d_x\delta\varphi_1 + ..., \quad \delta\vec{d}_2 = \vec{j}\,d_x\delta\varphi_2 + ..., \quad \delta\varphi_2 = -\alpha_2\delta\varphi_1 , \qquad (13)$$

and the components of $\vec{d}_1$ and $\vec{d}_2$ are determined by the expressions:

$$\vec{d}_1 = d\big(\sin(\chi_1 + \delta/2),\ 0,\ \cos(\chi_1 + \delta/2)\big),$$
$$\vec{d}_2 = d\big(-\sin\chi_2,\ 0,\ \cos\chi_2\big)$$

Taking into account that $\delta\varphi_1 \equiv \delta N_2$, we get the following final result:

$$<\delta\vec{d}_D^2>_2 = d^2 A_2(\chi_1,\chi_2)<\delta N_2^2>, \quad A_2(\chi_1,\chi_2) = \big(\sin(\chi_1+\delta/2)+a_2\sin\chi_2\big)^2. \qquad (14)$$

Analogously we get:

$$<\delta\vec{d}_D^2>_3 = d^2 A_3(\chi_1,\chi_2)<\delta N_3^2>, \quad A_3(\chi_1,\chi_2) = 1+a_3^2+2a_3\cos(\chi_1+\chi_2+\delta/2), \qquad (15)$$

$$<\delta\vec{d}_D^2>_4 = d^2 A_4(\chi_1,\chi_2)<\delta N_4^2>, \quad A_4(\chi_1,\chi_2) = 1+a_4^2+2a_4\cos(\chi_1+\chi_2+\delta/2), \qquad (16)$$

where $a_3 = (I_{y_1 y_1}^{(1)}/I_{y_1 y_1}^{(2)}) << 1$, and $a_4 = (I_{y_2 y_2}^{(2)}/I_{y_2 y_2}^{(1)}) << 1$.

The increment of $\delta\phi_1$ does not change the dipole moment of the left molecule. The compensating turn of the right molecule changes its dipole moment, however, this variation is negligibly small and we can ignore it. The analogous situation also takes place for the increment of $\delta\phi_2$.

With good accuracy the average value of $<\delta N_k^2>$ can be estimated according to classic formula:

$$<\delta N_k^2> = \frac{k_B T}{m_k \omega_k^2}, \quad k = 2,3,4, \qquad (17)$$

where $\omega_k^2 = k_{N_k N_k}/m_k$, $k = 1,...,6$, are the frequencies for normal vibrations and $k_{N_k N_k}$ are the elastic coefficients. In the more general case we can write:

$<\delta N_k^2> \approx \dfrac{k_B T}{m_k \omega_k^2} f(T/T_k)$ where the crossover function $f(T/T_k)$, $T_k = \hbar\omega_k/k_B$, is assumed to be similar to one determining the contribution of $k-$th vibration mode to the heat capacity (see (35) in the Section 8). The behavior of $f(T/T_k)$ for different vibration modes is presented in the Fig.2.

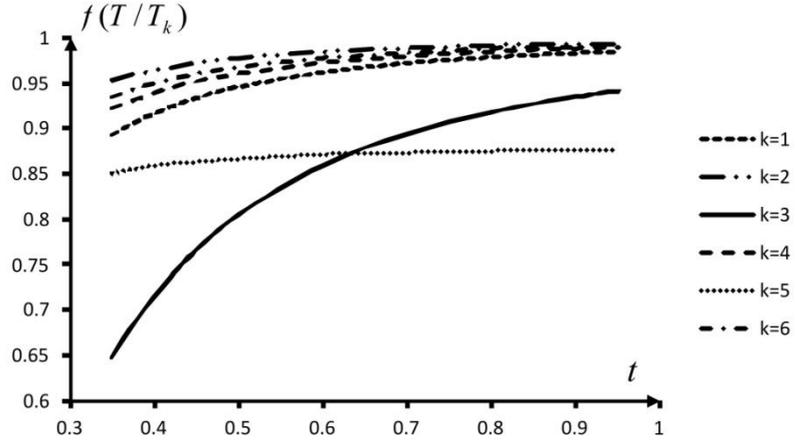

Fig.2. Temperature dependencies of the crossover functions $f(T/T_k)$ for different $k$.

Now let us define the effective polarizability $\alpha_d$ generated by the vibrations of a water dimer:

$$\alpha_d = \frac{<\delta\vec{d}_d^2>}{3k_BT}. \tag{18}$$

In accordance with the above we find for it:

$$\alpha_d = \alpha_d^{(2)} + \alpha_d^{(3)} + \alpha_d^{(4)}, \quad \alpha_d^{(k)} = \frac{d^2}{3mr_0^2\omega_0^2}\frac{A_k(\chi_1,\chi_2)}{m_k'\omega_k'^2}, \quad k = 2,3,4, \tag{19}$$

where $m_k' = m_k/(mr_0^2)$, $\omega_k' = \omega_k/\omega_0$, $\omega_0 = 10^{13}\,s^{-1}$. By order of magnitude:

$\left.\dfrac{d^2}{3mr_0^2\omega_0^2}\right/\alpha_e^{(m)} \approx 10^2$, where $\alpha_e^{(m)} \approx 1.05\cdot10^{-24}\,cm^3$ is the electronic polarizability of a water monomer, i.e. the effective polarizability of a dimer is essentially more than its electronic one ($\alpha_e^{(d)} \approx 2\alpha_e^{(m)}$).

Numerical values of the partial contributions: $\widetilde{A}_k(\chi_1,\chi_2) = \dfrac{A_k(\chi_1,\chi_2)}{m_k'\omega_k'^2}$, $k = 2,3,4$, to the effective polarizability (19) as well as other parameters are collected in the Table 1.

Table 1. Dimensionless frequencies, inertia moments and their ratios as well as partial contributions to the effective polarizability

|  | $N_1$ | $N_2$ | $N_3$ | $N_4$ | $N_5$ | $N_6$ |
|---|---|---|---|---|---|---|
| $\omega'_k$ , [18] | 3.52 | 2.28 | 6.97 | 2.97 | 10.64 | 2.71 |
| $m'_k$ |  | 1.449 | 2.025 | 0.602 | 1.326 | 1.327 |
| $a_{N_k}$ |  | 0.698 | 0.012 | 0.004 | 0.011 | 0.056 |
| $\widetilde{A}_k$ |  | 0.048 | 0.006 | 0.164 |  |  |

The relative values of the fluctuation contributions to the dimer dipole moment for different temperatures are presented in Table 2.

Table 2. The ratio $\sqrt{<\vec{\delta d}_d^2>}/d_d$ for different temperatures, calculated according to (19).

| $t$ | 0.45 | 0.5 | 0.55 | 0.6 | 0.65 | 0.7 | 075 | 0.8 | 0.85 | 0.9 | 0.95 |
|---|---|---|---|---|---|---|---|---|---|---|---|
| $\sqrt{<\vec{\delta d}_d^2>}/d_d$ | 0.128 | 0.134 | 0.141 | 0.147 | 0.153 | 0.159 | 0.165 | 0.170 | 0.175 | 0.180 | 0.185 |

As we can see the relative value of the dispersion for a dimer dipole moment does not exceed 8- 10% for all temperatures of liquid states.

## 4. Dielectric permittivity of a dimerized water vapor

The dielectric permittivity of a dimerized water vapor is determined by the expression [15,19]:

$$\frac{\varepsilon-1}{\varepsilon+2} = \frac{4\pi}{3}(n_m+n_d)\left[c_m(t)\left(\alpha_e^{(m)}+\frac{d_m^2}{3k_BT}\right)+c_d(t)\left(\alpha_e^{(d)}+\frac{\vec{d}_d^2+<\vec{\delta d}_d^2>}{3k_BT}\right)\right],$$

where $c_m = \dfrac{n_m}{n_m+n_d}$, $c_d = \dfrac{n_d}{n_m+n_d}$ are the numerical concentrations of monomers and dimers in water vapor, satisfying the standard normalization condition:

$$c_m+c_d=1. \tag{20}$$

Using (19) we can write

$$\frac{\varepsilon-1}{\varepsilon+2} = \frac{4\pi}{3}(n_m + n_d)\left[c_m\left(\alpha_e^{(m)} + \frac{d_m^2}{3k_BT}\right) + c_d\left(\alpha_e^{(d)} + \alpha_d + \frac{\vec{d}_d^2}{3k_BT}\right)\right]. \qquad (21)$$

Since the sum $n_m + n_d$ is not the quantity controlled experimentally, it should be connected with the mass density $\rho$ in accordance with (4). As a result we get

$$\frac{\varepsilon-1}{\varepsilon+2} = \frac{4\pi}{3}\frac{\rho/m_w}{1+c_d}\left[\alpha_e^{(m)} + \frac{d_m^2}{3k_BT} + c_d\left(\alpha_e^{(d)} - \alpha_e^{(m)} + \alpha_d + \frac{\vec{d}_d^2 - d_m^2}{3k_BT}\right)\right]. \quad (22)$$

If the dimer concentration is small the formula (22) reduces to

$$\frac{\varepsilon-1}{\varepsilon+2} \approx \frac{4\pi}{3}\frac{\rho}{m_w}\left[\alpha_e^{(m)} + \frac{d_m^2}{3k_BT} + c_d\left(\alpha_e^{(d)} - 2\alpha_e^{(m)} + \alpha_d + \frac{\vec{d}_D^2 - 2d_m^2}{3k_BT}\right) + ...\right]. \qquad (23)$$

## 5. Analysis of the dielectric permittivity of saturated water vapor

Let us determine the effective polarizability of a water molecule processing the experimental data on dielectric permittivity of a saturated water vapor according to the formula: $\alpha_{eff} = \frac{3}{4\pi}\frac{m_w}{\rho}\frac{\varepsilon-1}{\varepsilon+2}$. The corresponding temperature dependence of $\alpha_{eff}(t)$ is presented in Fig.3.

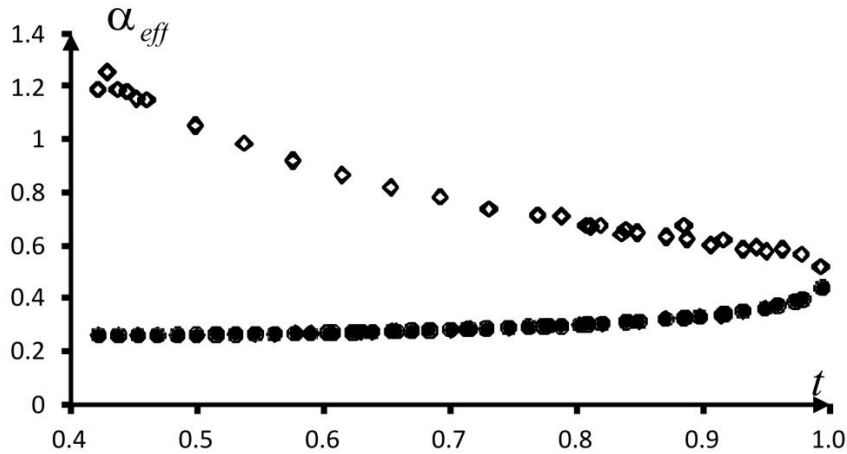

Fig.3. Temperature dependence of the effective polarizability for a water molecule: open rhombs correspond to saturated water vapor and dots – to liquid water.

*a) The effective polarizability and degree of dimerization outside the critical region*

The behavior of the product $\widetilde{\alpha}_{eff} \cdot t$, where $t = T/T_c$ is the dimensionless temperature ($T_c$ is the critical one) and $\widetilde{\alpha}_{eff} = \alpha_{eff}/r_{O_1 O_2}^3$ is the dimensionless effective polarizability, is presented in Fig.4.

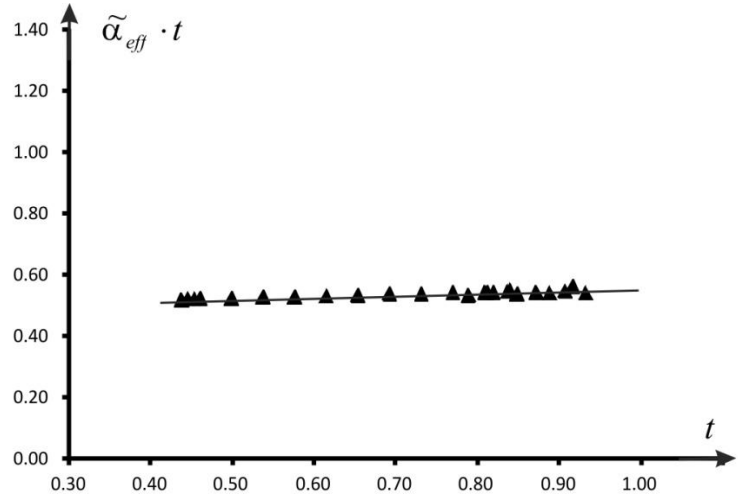

Fig.4. Temperature dependence of $\widetilde{\alpha}_{eff} \cdot t$ in the temperature interval $0.43 < t < 0.83$: triangles corresponds to experimental data, the solid line fits them with the help of least-square method.

As it follows from the Fig.4 the solid line is practically linear:

$\widetilde{\alpha}_{eff}(t) \cdot t = 0.0585 \cdot t + 0.4945$ within the temperature interval: $0.43 < t < 0.83$. The coefficient near $t$

$$\widetilde{\alpha}_{eff}(t) = 0.0585 \qquad (24)$$

is close to that for an isolated water molecule [1]: $\widetilde{\alpha}_e^{(is)} = 0.055$. Equating (24) to the expression:

$$\widetilde{\alpha}_{eff}(t) = \widetilde{\alpha}_e^{(m)} + c_d(t)\left(\widetilde{\alpha}_e^{(d)} - 2\widetilde{\alpha}_e^{(m)} + \widetilde{\alpha}_d(t)\right) + ..., \qquad (25)$$

following from (23), we conclude that the concentration of water dimers in a saturated water vapor within the temperature interval $0.43 < t < 0.83$ does not exceed 1%.

Note, that the small difference between $\widetilde{\alpha}_{eff}(t)$ and $\alpha_e^{(is)}$ for $0.43 < t < 0.83$ can be naturally explained with the help of two particle contributions to the effective polarizability of a molecule [15].

The contribution to $\tilde{\alpha}_{eff}(t)$, which is inverse proportional to $t$, should be identified with the combination of monomer and dimer dipole moments:

$$\frac{d_m^2}{3k_B T_c} + c_d(t)\frac{\vec{d}_D^2 - 2d_m^2}{3k_B T_c} = 13.132 \cdot r_{O_1 O_2}^3 .$$

In accordance with [15] the dipole moment of a dimer is equal to $\vec{d}_d^2 = 2d_m^2$. Therefore we get $d_m = 1.87\,D$ that practically coincides with the experimental value of dipole moment for a water monomer [1].

Thus, the noticeable dimerization in saturated water vapor is only observed in the temperature interval $0.85 < t < 1$ near the critical point. This conclusion is fully consistent with that in [20].

*b) The effective polarizability of a water molecule in the vicinity of the critical point*

Values of the combination $\tilde{\alpha}_{eff}(t) \cdot t$ for $0.85 < t < 1$ are presented in the Fig.5. Here the scattering of points is essentially larger than beyond the range of critical fluctuations, although it does not exceed ~4%.

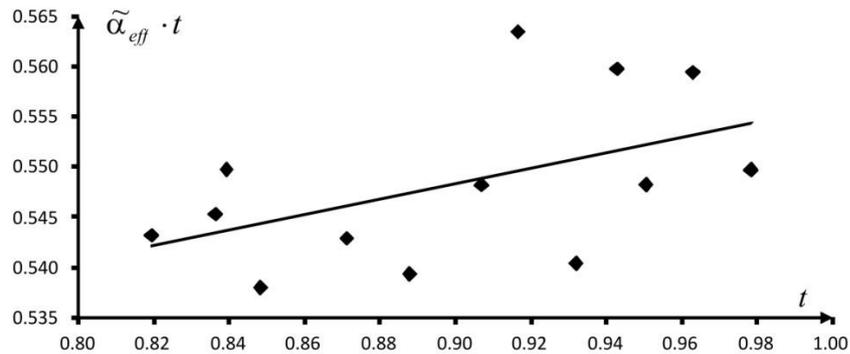

Fig.5.The combination $\tilde{\alpha}_{eff}(t) \cdot t$ vs. dimensionless temperature (solid line). The rhombuses correspond to experimental data for the dielectric permittivity from [21].

Values of the dimer concentration $c_d(t)$ can be calculate according to the equation:

$$\tilde{\alpha}_{eff}^{(exc)}(t) = \frac{1}{1 + c_d(t)}\left[\tilde{\alpha}_e^{(m)} + c_d(t)\left(\tilde{\alpha}_d(t) + \tilde{\alpha}_e^{(d)} - \tilde{\alpha}_e^{(m)}\right)\right], \qquad (26)$$

or

$$\tilde{d}_{eff}^2 = \frac{1}{1 + c_d(t)} \frac{\tilde{d}_m^2 + c_d(t)\left(\tilde{d}_d^2 - \tilde{d}_m^2\right)}{k_B T_c} . \qquad (27)$$

The numerical values of $\tilde{\alpha}_{eff}^{(exc)}(t)$ for a water molecule in different temperature intervals are presented in the first line of the Table 1. The values of $\tilde{\alpha}_d = 0.639$ and $\tilde{\alpha}_d = 0.577$ are calculated according to the definition (19) for different sets of frequencies from [18] and [16] correspondingly. In connection with this two series of values for $c_d(t)$ calculated according to (26) are given. Since these values of $c_d(t)$ are close further we will use the values of $c_d(t)$ from the second line. In fact, these values of $c_d(t)$ are the averages on the corresponding interval of temperature. These values of $c_d(t)$ are substituted to (27). The experimental and calculated values of the dipole combination $\tilde{d}_{eff}^2$ are given in the fourth and fifth lines. The difference between $\tilde{d}_{eff}^2(calc)$ and $\tilde{d}_{eff}^2(\exp)$ is connected with the scattering of experimental data for the dielectric permittivity of water vapor reaching 4% (see Fig.5).

Table 3. Comparative values of relevant parameters for different temperature intervals.

| $t$ | 0.8-0.98 | 0.83-0.98 | 0.85-0.98 |
|---|---|---|---|
| $\tilde{\alpha}_{eff}^{(exc)}(t)$ | 0.078 | 0.081 | 0.126 |
| $c_d$, [18] | 0.045 | 0.052 | 0.160 |
| $c_d$, [16] | 0.051 | 0.060 | 0.180 |
| $\tilde{d}_{eff}^2(\exp)$ | 0.479 | 0.475 | 0.433 |
| $\tilde{d}_{eff}^2(calc)$ | 0.493 | 0.493 | 0.492 |

It is necessary to note that the concentrations of dimers, determined here from the analysis of dielectric permittivity, quite satisfactory agree with those obtained from the analysis of the temperature dependence for the second virial coefficient $B(T)$ for a water vapor in [22] (see Table 4). Note that the last values of $c_d(t)$ are determined at some temperature, i.e. they are not averaged on temperature interval. Besides, our definition of the dipole concentration $c_d(t)$ is connected with $x_d$ from [22] by the formula: $c_d(t) = \dfrac{x_d}{2 - x_d}$.

<div align="center">

Table 4. Values of $c_d(t)$ according to [22].

| $t$ | 0.89 | 0.94 | 0.96 |
|---|---|---|---|
| $c_d(t)$ | 0.06 | 0.09 | 0.14 |

</div>

### *c) The effective polarizability of a molecule in liquid water*

Let us compare the effective polarizabilities of a water molecule in saturated vapor and liquid state. In the last case the behavior of $\alpha_{eff}(t) \cdot t$ is presented in the Fig.6.

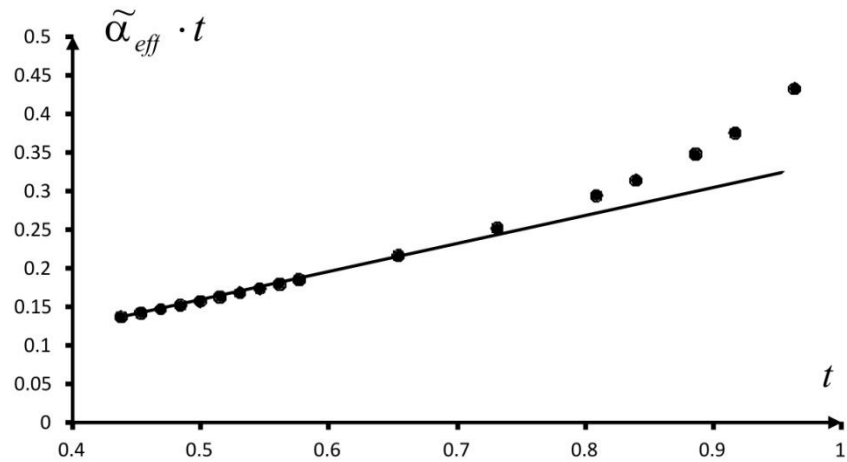

Fig.6. Temperature dependence of $\tilde{\alpha}_{eff} \cdot t$ for liquid branch of the coexistence curve within the interval: $t \in (0.42 \div 0.91)$. Circles correspond to experimental data [21], the solid line is the tangent to the curve $\alpha_{eff}(t) \cdot t$ at the triple point: $(\tilde{\alpha}_{eff} \cdot t)_{\tan g} = 0.302 \cdot t - 0.018$.

In order to interpret the behavior of $\tilde{\alpha}_{eff}$ we can represent the effective polarizability in the view:

$$\tilde{\alpha}_{eff}(t) = \tilde{\alpha}_{eff}^{(exc)}(t) + \frac{\tilde{d}_{eff}^{2}(t)}{3t}, \qquad (28)$$

where $\tilde{\alpha}_{eff}^{(exc)}(t) = \tilde{\alpha}_d(t) + ...$ is the contribution arising due to vibrational excitations of clusters and $\tilde{d}_{eff}(t) = \dfrac{d_{eff}(t)}{\sqrt{k_B T_c r_{OO}^3}}$.

Near the triple point the effective polarizability of a molecule tends to $\tilde{\alpha}_{eff}^{(exp)} \approx 0.302$ and $\tilde{d}_{eff}^{(exp)}(t) \approx 0$. This value of $\tilde{\alpha}_{eff}^{(exp)}$ approximately corresponds to the half of dimensionless effective polarizability $\tilde{\alpha}_d(t)$ of a dimer. Since the effective dipole moment is practically equal to zero, we should conclude that 1) the observed value of the effective polarizability per molecule is generated by the set of dimers; 2) two neighboring dimers are in antiparallel states similar to those for tetramer; 3) three dimers can form hexamers similarly to that in hexagonal ice. In these and similar cases dipole moments are equal to zero. In liquid state water dimers are the most stable among all enumerated clusters, though all they are short-living. So, we assume that dielectric properties of liquid water can be considered as those for an ensemble of dimers. The first steps in this direction had been made in [23-26].

In accordance with this we expect that near the triple point ($t < 0.6$):

$$\alpha_{eff}^{(exp)}(t) \approx \frac{1}{2}\alpha_d(t). \qquad (29)$$

For $t > 0.6$ the situation becomes more complicated. The effective dipole moment per molecule $d_{eff}(t)$ becomes different from zero. It is naturally explained by fluctuations of angle between two neighboring dimers (forming antiparallel states near the triple point). We want to devote a separate paper to this consideration.

## 6. The contribution of attractive forces to the heat-capacity of a water vapor

In this Section the main attention is focused on the temperature dependence of the heat capacity of water vapor. Its behavior, corresponding to experimental data [27-26], is presented in the Fig.7. The heat capacity is represented in dimensionless units: $i_Q(t) = 2C_V / k_B N_A$. These units we can interpret as the number $i_Q$ of thermal degrees of freedom per molecule. Near the triple point $i_Q \approx 6$, that corresponds to 3 translational and 3 rotational degrees of freedom. If temperature increases the following contributions: 1) the interaction between particles; 2) dimerization of water molecules and 3) the excitation of dimer vibrations, influence on the behavior of the heat capacity. The dimerization diminishes the number of particles so $i_Q$ will decrease. However, the excitation of dimer vibrations will lead to the increase of the heat capacity. The formation of tetramers and multimers of higher order will lead to further changes of the heat capacity.

Let us consider all these contributions separately.

*a) Influence of the intermolecular interaction on the temperature dependence of the heat capacity*

Following [28] the free energy of the dense enough vapor per molecule is determined by the expression:

$$f(T,n) = f_{id} + nTB(T) + \ldots,$$

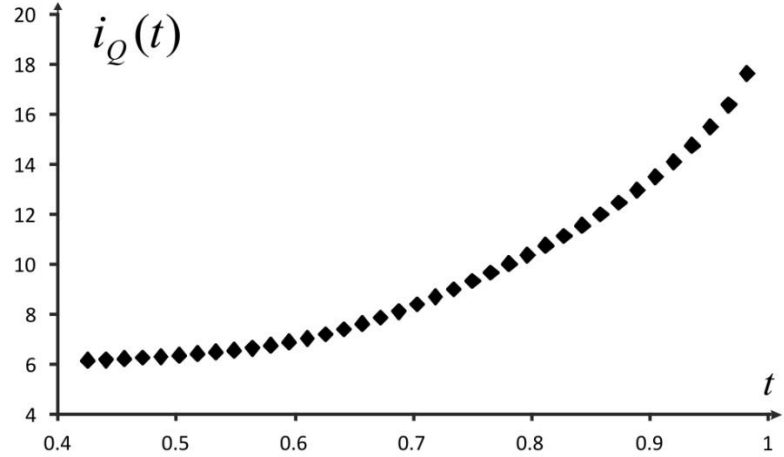

Fig.7. Temperature dependence of the number $i_Q$ of thermal degrees of freedom for saturated water vapor.

where

$$B(T) = \frac{1}{2}\int \left(1 - \exp\left(-\beta U_a(r)\right)\right)dV , \qquad (30)$$

and $U_a(r)$ is the averaged interparticle potential. The last is determined by the equation:

$$\frac{1}{\Omega^2}\int_\Omega\int_\Omega \exp\left(-\beta U(r,\Omega_1,\Omega_2)\right)d\Omega_1 d\Omega_2 = \exp\left(-\beta U_a(r)\right), \qquad (31)$$

where $U(r,\Omega_1,\Omega_2)$ is the bare potential of the interparticle interaction between water molecules of type SPC and the others ($d\Omega_i$, $i=1,2$ denote the elementary volumes corresponding to angular variables, $\Omega$ is the full volume for angular variables). This definition of the averaged potential allows us to conserve the value of the configuration integral in two particle approximation:

$$\frac{1}{\Omega^2}\int_V\int_\Omega\int_\Omega\left(1-\exp\left(-\beta U(r,\Omega_1,\Omega_2)\right)\right)d\Omega_1 d\Omega_2 dV = \int\left(1-\exp\left(-\beta U_a(r)\right)\right)dV .$$

The heat capacity at constant volume is determined in the standard way and it is equal to

$$c_\upsilon(T) = c_\upsilon^{(id)} + c_\upsilon^{(\text{int})}, \quad c_\upsilon^{(id)} = 6, \quad c_\upsilon^{(\text{int})} = -nT\left(2B'(T) + TB''(T)\right)\big|_\upsilon .$$

With suitable accuracy the averaged potential $U_a(r)$ can be approximated by the Lennard-Jones potential. Then the heat capacity per molecule on the coexistence curve equals to:

$$c_\upsilon^{(\text{int})}(t) = \frac{2\pi}{3}k_B\tilde{n}_{cc}(t)\left(4\tilde{\varepsilon}_a/t\right)^2\gamma(\tilde{\varepsilon}_a/t),$$

$$\gamma(z) = \int_0^\infty e^{-4z[(1/x)^4-(1/x)^2]}[(1/x)^4-(1/x)^2]^2 dx \tag{32}$$

where the dimensionless variables are determined according to: $t = T/T_c$, $\tilde{n}_{cc}(t) = n_{cc}(t)\sigma^3$ and $\tilde{\varepsilon}_a = \varepsilon_a/k_B T_c$. This expression for $c_\upsilon^{(\text{int})}(\tilde{n},t)$ is correct until the concentration of dimers is negligibly small, i.e. for $0.42 < t < 0.83$ [22].

### b) Influence of the monomer-dimer and dimer-dimer interactions on the temperature dependence of the heat capacity

The free energy of a dense enough water vapor consisting of monomers and not excited dimers is determined by the expression:

$$f = f_{id} + T\left[c_m^2 B_{mm}(T) + 2c_m c_d B_{md}(T) + c_d^2 B_{dd}(T)\right] + ..., \tag{33}$$

where

$$B_{ik}(T) = \frac{1}{2}\int\left(1-\exp\left(-\beta U_{ik}^{(a)}(r)\right)\right)dV, \quad i,k = m,d .$$

All intermediate calculations are similar to those in the previous Section. So, the final contribution of interparticle interactions to the heat capacity per molecule takes the following form:

$$c_v^{(\text{int})}(t) = \frac{2\pi}{3} \frac{k_B}{(1+c_d)^2} \frac{\rho_{cc}(t)}{m_w} \left[(1-c_d)^2 \zeta_m(t) + 2c_d(1-c_d)\zeta_{md}(t) + c_d^2 \zeta_d(t)\right]$$

$$\zeta_j = \sigma_j^3 (4\tilde{\varepsilon}_j^{(a)}/t)^2 \gamma\left(\tilde{\varepsilon}_j^{(a)}/t\right), \quad j = m, md, d.$$

It is assumed that the bare interaction between the water monomers is described by known potentials of the SPC-type. The averaged potentials are calculated by formula (31) and approximated by the Lennard-Jones potential. The corresponding values of its parameters $\tilde{\varepsilon}_m^{(a)}$ and $\sigma_m$ for water monomers are presented in the Table 5.

Table 5. The values of $\sigma_m$ and $\tilde{\varepsilon}_m^{(a)}$ for averaged potentials of different bare intermolecular potentials.

|  | $\sigma_m, A$ | $\tilde{\varepsilon}_m^{(a)}$ |
|---|---|---|
| SPC | 2.62 | 2.95 |
| TIP3P | 2.467 | 2.792 |
| TIP4P | 2.476 | 2.624 |
| TIP5P | 2.458 | 2.467 |
| MP | 2.80 | 2.187 |

We note that screening effects are only taken into account for the MP [ 29].

The parameters of the Lennard-Jones potential corresponding to the dipole-dipole interaction of two dimers are determined in Appendix 1. The corresponding parameters for the averaged potentials describing the monomer-dimer interaction are determined by approximate formulas [30]:

$$\varepsilon_{md}^{(a)} = \sqrt{\varepsilon_m^{(a)} \varepsilon_d^{(a)}} \text{ and } \sigma_{md} = \frac{1}{2}(\sigma_m + \sigma_d) \ . \tag{34}$$

The relation $\dfrac{n_m + n_d}{\rho/m_w} = \dfrac{1}{1+c_d}$, following from (4), is also used.

## 7. Vibration contribution to the heat capacity

The vibration contribution to the free energy per water dimer is determined by the standard expression [28]:

$$F_{vib} = -T \ln Z_{vib}, \quad Z_{vib} = \sum_{k=1}^{6} \sum_{n=0}^{N} \exp\left(-E_n^{(k)}/T\right),$$

where

$$E_n^{(k)} = -\varepsilon_0^{(k)} + \hbar\omega_k(n+1/2),$$

$-\varepsilon_0^{(k)}$ is the ground state energy for $k-th$ mode of a dimer, $N$ is the maximal number of vibration levels in the potential well of the finite depth. Here we suppose that the potential well can be approximated by the parabola with satisfactory accuracy. The full number of thermal degrees of freedom equals to:

$$i_Q^{(\upsilon)}(t) = \frac{c_d}{1+c_d} \sum_{k=1}^{6} i_Q^{(k)}(t) \qquad (35)$$

where

$$i_Q^{(k)}(t) = 2f(T/T_k), \quad f(T/T_k) = \frac{t_k^2}{t^2} \frac{\exp(t_k/t)}{\left(\exp(t_k/t)-1\right)^2}, \qquad (36)$$

$t_k = T_k/T_c$ and $\dfrac{n_d}{\rho/m_w} = \dfrac{c_d}{1+c_d}$ in accordance with (4).

### 8. Analysis of different contributions to the heat capacity of a water vapor

In this Section we consider consequently: 1) the contribution $i_Q^{(mm)}(t)$ of monomer-monomer interactions to the number of the thermal degrees of freedom for a water vapor; 2) the corrections $i_Q^{(md)}(t)$ and $i_Q^{(dd)}(t)$ caused by monomer-dimer and dimer-dimer interactions and 3) the contribution $i_Q^{(\upsilon)}(t)$ caused by vibration modes of dimers. In order to compare the calculated values of

$$i_Q(t) = 6 + i_Q^{(mm)}(t) + i_Q^{(md)}(t) + i_Q^{(dd)}(t) + i_Q^{(\upsilon)}(t)$$

with experimental data we will use values of the molar dimer concentration $c_d(t)$ from [12] determined with the help of the second virial coefficient.

### a) The contribution $i_Q^{(mm)}(t)$ of monomer-monomer interactions

We start from the expression:

$$i_Q(t) = 6 + i_Q^{(mm)}(t) + ...,$$

where

$$i_Q^{(mm)} \approx \frac{2\pi}{3}\, \tilde{\rho}_{cc}(t)\left(4\tilde{\varepsilon}_m^{(a)}/t\right)^2 \gamma\left(\tilde{\varepsilon}_m^{(a)}/t\right)+...$$

and $\tilde{\rho}_{cc}(t) = \dfrac{\rho_{cc}(t)\sigma_m^3}{m_w}$ . Note that numerical values of $\tilde{\varepsilon}_m^{(a)}(t)$ and $\sigma_m(t)$ for the SPC-potential, calculated in [22, 29, 31], are dependent on temperature (see Fig.8), though this effect becomes to be essential only in the narrow vicinity of the critical point.

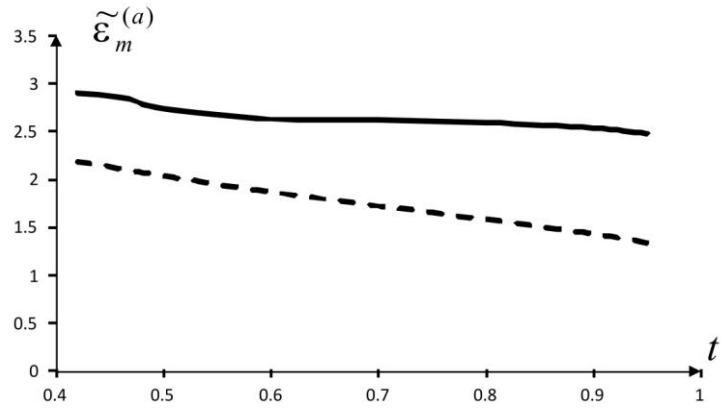

Fig.8. Temperature dependence of $\tilde{\varepsilon}_m^{(a)}$ calculated in [29,31] for the averaged SPC-potential (solid line) and MP (dotted line).

The comparison with experimental data for these potentials are presented in the Fig.9.

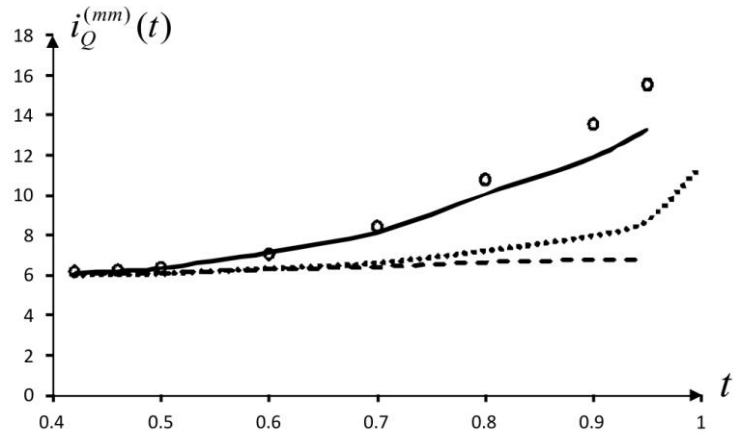

Fig.9. Temperature dependencies of the numbers of thermal degrees of freedom: open circles represent experimental data [27], the solid line corresponds to $i_Q^{(mm)}(t)$ for the averaged SPC-potential, dashed line - to the averaged MP and the dotted line – to the experimental values of the second virial coefficient [32].

Here, the behavior of the second virial coefficient is reproduced essentially better by the averaged MP, while the combination (31) of the first and second temperature derivatives reproduces better the behavior of the heat capacity with the averaged SPC-potential. The last result is occasional, although the effective SPC-potential can partly reflect the contributions $i_Q^{(md)}(t)$ and $i_Q^{(dd)}(t)$ of monomer-dimer and dimer-dimer interactions. The last are described by the formula (33). At that, some fine details in temperature dependences of $i_Q^{(md)}(t)$ and $i_Q^{(dd)}(t)$ are connected with the temperature dependence of $\tilde{\varepsilon}_m^{(a)}$ shown in the Fig.8.

The considerable deviation of the dashed and dotted lines from the experimental circles for $t > 0.6$ in the Fig.9 explicitly testifies about the important role of thermal excitations of dimers since the monomer-dimer and dimer-dimer interactions cannot influence on the heat capacity more than monomer-monomer interactions.

### b) Role of the dimer contributions

In order to reproduce the temperature dependence of $i_Q(t)$ we will use values of $c_d(t)$ determined in [22]. The corresponding results are presented in the Fig.10.

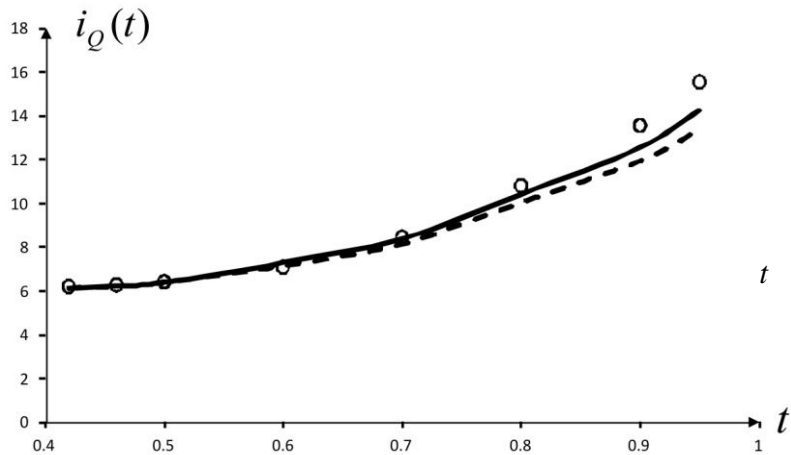

Fig.10. Temperature dependencies of $i_Q(t)$ in different approximations (experimental data are given by open circles): dotted line corresponds to $i_Q(t) = 6 + i_Q^{(mm)}(t) + i_Q^{(md)}(t) + i_Q^{(dd)}(t)$, solid line $- i_Q(t) = 6 + i_Q^{(mm)}(t) + i_Q^{(md)}(t) + i_Q^{(dd)}(t) + i_Q^{(v)}(t)$.

We see that the experimental data are reproduced with good accuracy up to 0.8. In this case, 1) the contributions caused by monomer-dimer and dimer-dimer interactions increase with temperatures and they achieve about 1% and 2) the contribution of dimer vibrations also increases with temperature and gives approximately 6% . The use of $c_d(t)$ from other works [ 9,33,34] leads to the worse agreement between experimental and calculated data. The absence of the full agreement in the interval $0.8 < t < 0.95$ can be explained by 1) the uncertainty in the determination of $c_d(t)$ and 2) the necessity to take into account trimers and clusters of higher order [23, 35].

### Discussion of the results obtained

In the work we consider the physical nature of the effective polarizability per molecule and heat capacity for water vapor as well as their temperature dependencies. If we take into account the dimerization and the thermal excitations of dimers it allows us to reproduce the behavior of the effective polarizability and heat capacity with quite satisfactory accuracy. This circumstance is really very important because the experimental data for the heat capacity of water vapor are measured with high accuracy. From our point of view the reproduction of temperature dependencies for the effective polarizability and heat capacity for saturated water vapor is one of the best tests for different estimates of $c_d(t)$. The values of $c_d(t)$ in [22] were calculated with the help of experimental data from [32] for the second virial coefficient. This test allows us to conclude that they are close to optimal ones. The close results can be also obtained on the base of experimental material from [36] (Fig.11).

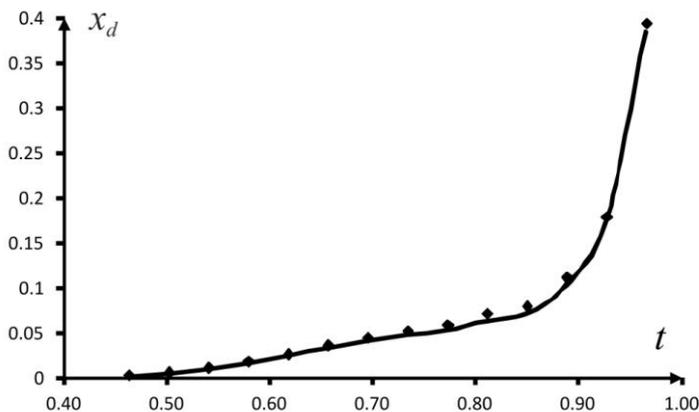

Fig.11. Temperature dependencies for $x_d(t)$ obtained in [22] with the help of experimental data [32, 36] for the second virial coefficient of the saturated water vapor.

Note, we do not use the notion of the H-bond. All properties of dimers are studied within the electrostatic representations. In particular, we describe fluctuations of dipole moment for a dimer as well as equilibrium parameters of the latter. The analogous situation also takes place for dimer vibrations. However, this question is not considered in our work, as we have used the known experimental values for vibration frequencies. We would like to stress that the behavior of the effective polarizability and heat capacity also have been investigated in [23, 31]. However, in [23, 31] we used the representations based on the H-bond vibrations. In this case, the consideration of many details was impossible. Though, if we use the assumption about the electrostatic nature of the H-bond [17,37-41], both these considerations will become qualitatively equivalent.

The generalization of our approach to liquid water is possible and it seems to us very important. In this case we face with a complex problem to describe correctly the clusterization in liquid water. However, our assumption about the possibility to model electrostatic properties of multimers with the help of a set of dimers allows us to construct the satisfactory picture of liquid water properties. Within this picture the important role will belong to the number of H-bonds per molecule, treated as electrostatic objects. Within similar picture the minimum of the heat capacity at constant pressure, observed in [42,43] near $36^0 C$ , is reproduced satisfactorily. It is very important for the physics of alive. We plan to consider this question in a separate paper.

**Appendix 1.** The Lennard-Jones potential to approximate optimally the behavior of the averaged potential for dipole-dipole interactions.

Let us consider this problem for the bare potential $\Phi_{dd}(1,2)$ of dipole-dipole interaction between water dimer. We start from the following definition of the averaged potential $U_a(r)$ (details and designations see in [44]):

$$\exp\left(-\beta U_a(r)\right) = \oint_{\Omega_1=4\pi} \frac{d\Omega_1}{4\pi} \oint_{\Omega_2=4\pi} \frac{d\Omega_2}{4\pi} \exp\left(-\beta\Phi(1,2)\right) \qquad (A1)$$

The numerical values of $U_a(r)$ calculated by (A1) for some set of point is presented in Fig.A1 (open circles).

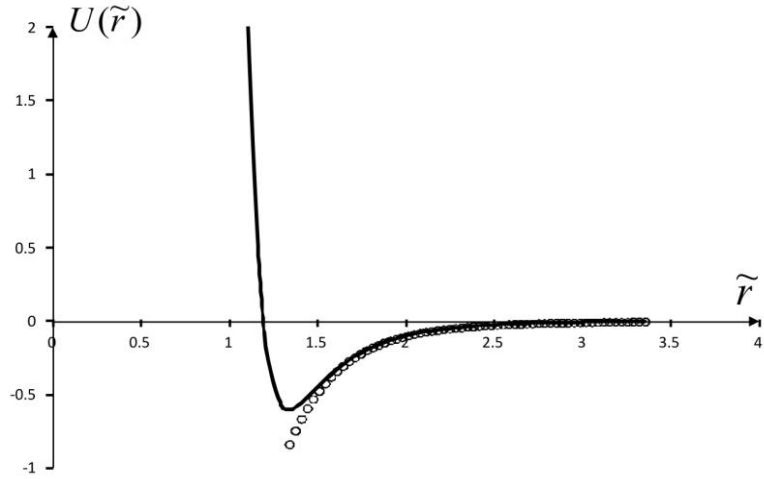

Fig.12. The comparative behavior of the averaged potential $U_a^{(dd)}(\tilde{r})$ (open circles) which corresponds to dipole-dipole interaction of two dimers and the suitable Lennard-Jones potential (solid line).

In order to find the parameters $\tilde{\varepsilon} = \varepsilon / k_B T_c$ and $\tilde{\sigma} = \sigma / r_{OO}$ for the suitable Lennard-Jones potential:

$$U_{LJ}(\tilde{r}) = 4\tilde{\varepsilon}\left[\left(\frac{\tilde{\sigma}}{\tilde{r}}\right)^n - \left(\frac{\tilde{\sigma}}{\tilde{r}}\right)^6\right]$$

where $\tilde{U}_a = U_a / k_B T_c$, we suppose that 1) numerical values of $U_a(\tilde{r}) \cdot \tilde{r}^n$ are fitted by the combination

$$L(\tilde{r}) = -4\tilde{\varepsilon}\tilde{\sigma}^6 \cdot \tilde{r}^{n-6} + 4\tilde{\varepsilon}\tilde{\sigma}^n \equiv -a \cdot \tilde{r}^{n-6} + b,$$

considered as a linear function of $\tilde{r}^{n-6}$ and 2) optimal value of $n$ correspond to that case for which points $U_a(\tilde{r}) \cdot \tilde{r}^n$ are maximally fitted to the line $L(\tilde{r}) = -a \cdot \tilde{r}^{n-6} + b$ for all $\tilde{r}$.

The parameters $\tilde{\varepsilon}$ and $\tilde{\sigma}$ are connected with $a$ and $b$ by the relations: $\tilde{\sigma} = (b/a)^{1/(n-6)}$, $\tilde{\varepsilon} = a^2/(4b) \cdot$ One can verify that the optimal value of $n$ for dipole-dipole interaction equals to $n = 12$. The parameters $\tilde{\varepsilon}$ and $\tilde{\sigma}$ takes the following values:

$$\tilde{\varepsilon} = 0.61, \quad \tilde{\sigma} = 1.19.$$

The degree of fitting $L(\tilde{r}) = -a \cdot \tilde{r}^6 + b$ to $U_a(\tilde{r}) \cdot \tilde{r}^{12}$ is presented in the Fig.12.

It is necessary to note that 1) $n = 12$ guarantees the similarity of potentials for water and argon and it leads to argon-like temperature dependencies of the specific volume and evaporation heat for water [31] and 2) the parameters $\tilde{\varepsilon}$ and $\tilde{\sigma}$ are weak functions of temperature.

## Literature

1. D.Eisenberg and V. Kauzmann, *The Structure and Properties of Water* (Oxford

   University Press,New York,USA, 1969, p 310 ).

2. K.Burrows, E.R.Pike, J.M. Vaughan. Nature **260**,131(1976). doi:10.1038/260131a0

3. G. E. Ashwell, P. A. Eggett, R. Emery, H.A.Gebbie. Nature **247**,196 (1974).

doi:10.1038/247196a0

4. L. A. Curtiss, D. J. Frurip, M. J. Blander. Chem. Phys.**71**, 2703 (1979). doi:10.1063/1.438628

5. R. A. Bohlander, H. A. Gebbie, G. W. F. Pardoe. Nature **228**, 156 (1970). doi:10.1038/228156a0

6.  J. Hargrove. Atmos. Chem. Phys. Discuss.**7**, 11123 (2007).

7. A. J. L. Shillings, S. M. Ball, M. J. Barber, J. Tennyson, R. L. Jones. Atmos. Chem.

Phys.**11**,4273 (2011). doi:10.5194/acp-11-4273-2011

8. A.A.Vigasin. J. Quant. Spectr. & Rad.Transf. **64**, 25(2000). doi:10.1016/S0022-4073(98)00142-3

9. A.A.Vigasin, A.I. Pavlyuchko, Y. Jin, S. Ikawa.  J. of Mol. Str. 742,173 (2005).

doi.org/10.1016/j.molstruc.2004.12.060

10. C. J. Leforestier. Chem. Phys.**140**, 074106 (2014). doi: 10.1063/1.4865339

11. J. O. Hirschfelder, F. T. McClure, I. F. Weeks. J. Chem. Phys **10**, 201 (1942).

doi.org/10.1063/1.1723708

12. D. Stogrynt, J. O. Hirschfelder J. Chem. Phys. **31,** 6, 1531 (1959). doi.org/10.1063/1.1730649

13. G. N.I. Clark, D. C.Christopher, J. D. Smith, R. J. Saykally. Molecular Physics **108,** 11, 1415

(2010). doi.org/10.1080/00268971003762134

14. Y. Scribano, N. Goldman, R. J. Saykally. J. Phys. Chem. A **110**, 5411 (2006).

DOI: 10.1021/jp056759k

15. N.P. Malomuzh, V.N. Makhlaichuk, S.V. Hrapatiy. Russian Journal of Physical Chemistry A **88**, No. 8, 1431 (2014). DOI: 10.1134/S0036024414080172

16. J.R.Reimers, R.O.Watts. Chemical Physics **85**, 83 (1984),. doi:10.1016/S0301-0104(84)85175-7

17. H. J. C. Berendsen, J. P. M. Postma, W. F. van Gunsteren and J. Hermans, in B. Pullman (ed.), Intermolecular Forces (Reidel, Dordrecht, 1981) p331.

18. M.J.Smit, G.C.Groenenboom, P.E.S.Wormer, Ad van der Avoird, R.Bukowski, K.Szalewicz. J. Phys. Chem. A **105**, 6212 (2001). DOI: 10.1021/jp004609

19. H. Fröhlich, *Theory of Dielectrics: Dielectric Constant and Dielectric Loss* (Clarendon, Oxford, 1958).

20. V.L.Kulinskii, N.P.Malomuzh. Phys.Rev.E **67**, 011501 (2003). doi.org/10.1103/PhysRevE.67.011501

21. D.P. Fernandez, Y. Mulev, A. R. H. Goodwin, J. M. H. Levelt Sengers. J. Phys. Chem. Ref. Data **24**, 1 (1995). http://dx.doi.org/10.1063/1.555977

22. N. P. Malomuzh, V. N. Mahlaichuk, S. V. Hrapatiy. Russian Journal of Physical Chemistry A **88**, No. 8, 1287 (2014). DOI: 10.1134/S003602441406017X

23. N. P. Malomuzh , V. N. Makhlaichuk , P. V. Makhlaichuk , K. N. Pankratov. Journal of Structural Chemistry., **54**, Supplement 2, 205 (2013). DOI 10.1134/S0022476613080039

24. A.I. Fisenko, N.P. Malomuzh, A.V. Oleynik. Chem. Phys. Lett. **450**, 297 (2008). http://dx.doi.org/10.1016/j.cplett.2007.11.036

25. L.A. Bulavin, A.I. Fisenko, N.P. Malomuzh Chem. Phys. Lett.**453,** 183 *(2008).* http://dx.doi.org/10.1016/j.cplett.2008.01.028

26. L.A. Bulavin, T.V. Lokotosh, N.P. Malomuzh J. Mol. Liquid **137**, 1 (2008). http://dx.doi.org/10.1016/j.molliq.2007.05.003

27. E.W. Lemmon, M.O. McLinden and D.G. Friend, "Thermophysical Properties of Fluid Systems" in NIST Chemistry WebBook, NIST Standard Reference Database Number 69, Eds. P.J. Linstrom and W.G. Mallard, National Institute of Standards and Technology, Gaithersburg MD, 20899, http://webbook.nist.gov.

28. L. D. Landau and E. M. Lifshitz, Course of Theoretical Physics, Vol. 5: *Statistical Physics* (Nauka, Moscow,121995; Pergamon, Oxford, 1980).

29. M.V. Timofeev.  Ukr. J. Phys., **61**, N 10,893(2016). http://dx.doi.org/10.15407/ujpe61.10.0893

30. J.O. Hirschfelder, Ch.F. Curtiss, R.B. Bird. Molecular theory of gases and liquids. John Wiley and Sons, inc., New York, 1954.

31. S. V. Lishchuk, N. P. Malomuzh, P. V. Mahlaichuk, Phys. Lett. A **374**, 2084 (2010).

32. A. H. Harvey and E. W. Lemmon. J. Phys. Chem. Ref.Data 33, 369 (2004).

http://dx.doi.org/10.1063/1.1587731

33. G. T. Evans and V. Vaida. J. Chem. Phys. **113**, 6652 (2000).

http://dx.doi.org/10.1063/1.1310601

34. Y. Scribano, N. Goldman, R. J. Saykally, C. Leforestier J. Phys. Chem. A **110**, 5411 (2006). DOI: 10.1021/jp056759k

35. M. Yu. Tretyakov and D. S. Makarov. The Journal of Chemical Physics **134**, 084306 (2011); doi: http://dx.doi.org/10.1063/1.3556606.

36. Moscow Power Engineering Institute, Mathcad Calculation Server.

http://twt.mpei.ac.ru/MCS/Work_sheets/PhyRefBook/13_2.xmcd

37. N. D. Sokolov  Uspekhi Fizicheskikh Nauk **57**, 205 (1955)

 DOI: 10.3367/UFNr.0057.195510d.0205

38. W.L. J. Jorgensen Am. Chem. Soc.**103**, 335 (1981). DOI: 10.1021/ja00392a016

39. M. Dolgushin, V. Pinchuk, preprint ITP-76-49R, (Inst. for Theor. Phys. of the NASU, Kiev, 1976) (in Russian).

40. I.V. Zhyganiuk, M.P. Malomuzh Ukr. J. Phys. **60**, 960 (2015). doi:10.15407/ujpe60.09.0960

41. P.V. Makhlaichuk, M.P. Malomuzh, I.V. Zhyganiuk. Ukr. J. Phys., **58**, 278 (2013).

42. R. C. Dougherty and L. N. Howard. J. Chem. Phys. 109, 7379(1998).

43. National Institute of Standards and Technology, A gateway to the data collections. Available at http://webbook.nist.gov

44. Pavlo V. Makhlaichuk, Victor N. Makhlaichuk, Nikolay P. Malomuzh. Journal of Molecular Liquids, **225**, 577 (2017). https://doi.org/10.1016/j.molliq.2016.11.101.